\documentclass[twocolumn,showpacs,preprintnumbers,amsmath,amssymb, prb]{revtex4}
\usepackage{graphicx}
\usepackage{dcolumn}
\usepackage{bm}
\usepackage{wasysym}
\usepackage{textcomp}
\usepackage{epstopdf}
\usepackage{stmaryrd}
\usepackage{txfonts}

\begin{document}

\title{Anomalous Magnetic Properties of Sr$_{2}$YRuO$_{6}$}

\author{Ravi P. Singh and C. V. Tomy}

\affiliation{Department of Physics, Indian Institute of Technology Bombay, Mumbai
400 076, INDIA}
\begin{abstract}
Anomalous magnetic properties of the double perovskite ruthenate compound
Sr$_{2}$YRuO$_{6}$ are reported here. Magnetization measurements
as a function of temperature in low magnetic fields show clear evidence
for two components of magnetic order ($T_{M1}\sim32$~K and $T_{M2}\sim27$~K)
aligned opposite to each other with respect to the magnetic field
direction even though only Ru$^{5+}$ moments can order magnetically
in this compound. The second component of the magnetic order at $T_{M2}\sim27$~K
results only in a magnetization reversal, and not in the negative
magnetization when the magnetization is measured in the field cooled
(FC) mode. Isothermal magnetization ($M$-$H$) measurements show
hysteresis with maximum coercivity ($H_{c}$) and remnant magnetization
($M_{r}$) at $T\sim$27~K, corroborating the presence of the two
oppositely aligned magnetic moments, each with a ferromagnetic component.
The two components of magnetic ordering are further confirmed by the
double peak structure in the heat capacity measurements. These anomalous
properties have significance to some of the earlier results obtained
for the Cu-substituted superconducting Sr$_{2}$YRu$_{1-x}$Cu$_{x}$O$_{6}$
compounds.
\end{abstract}


\maketitle

\section{INTRODUCTION}

Sr$_{2}$YRuO$_{6}$ belongs to the family of double perovskite insulators,
Sr$_{2}Ln$RuO$_{6}$ ($Ln$ = rare earth or Y) \cite{P. C. Donohue (1977)},
where the Ru ions exist in the pentavalent state (Ru$^{5+}$) with
a high-spin state $^{4}A_{2g}$ and $4d^{3}$ configuration ($J=3/2$)\textbf{.}
Even though the structure of these compounds can be derived from the
well known perovskite structure of SrRuO$_{3}$ by replacing alternate
Ru ions with $Ln$ ions \cite{P. D. Battle and W. J. Macklin (1984)},
these compounds do not show any similarity to their parent compound
SrRuO$_{3}$, which is a ferromagnetic metal. The layered structure,
consisting of alternate $Ln$RuO$_{4}$ and SrO planes, accommodates
both the Ru and rare earth atoms in the same $Ln$RuO$_{4}$ plane
and hence both the atoms share the same site symmetry ($B$-site of
the perovskite structure ABO$_{3}$). The alternating positions of
the Ru and $Ln$ atoms in the unit cell result in two type of interactions
between the Ru atoms; (i) direct interaction of Ru-O-O-Ru and (ii)
indirect interaction through the rare earth atoms, Ru-O-$Ln$-O-Ru.
Since the compounds having nonmagnetic $Ln$ ions (Y and Lu) are also
found to order magnetically \cite{P. D. Battle and W. J. Macklin (1984),P. D. Battle and C.W. Jones (1989)},
the direct interaction is assumed to be stronger than the indirect
interaction through the rare earth atoms. Among the Sr$_{2}Ln$RuO$_{6}$
compounds, Sr$_{2}$YRuO$_{6}$ has captured additional interest due
to the occurrence of superconductivity when Ru is partially ($\leq$
15\%) replaced by Cu \cite{Wu (2000),Marco (2000),Blackstead (2000),D. R. Harshman (2000),H. A. Blackstea (2001),Harshman  (2003)}.
Cu is found to get substituted at the Ru site in the YRuO$_{4}$ planes
and thus the structure of the substituted compounds remains the same
as that of the parent compound, without creating any additional Cu-O
planes \cite{H. A. Blackstea (2001)}.

The parent compound Sr$_{2}$YRuO$_{6}$ is known to be an antiferromagnetic
insulator with the Ru moments ordering at $T_{N}=26$~K \cite{P. D. Battle and W. J. Macklin (1984)}.
The magnetic ordering temperature ($T_{N}$) was inferred as 26~K
from the position of the peak in the magnetization measurements. Neutron
diffraction measurements at 4.2~K have confirmed the magnetic ordering
of the Ru moments, consisting of a type I AFM structure. Due to the
monoclinic distortion of the structure, the compound is expected to
show canting of the Ru moments resulting from the Dzyaloshinsky-Moriya
(D-M) interactions \cite{Dzyaloshinsky  (1958),T. Moriya  (1960)}
among the antiferromagnetically ordered spins. How such a compound
becomes a metallic magnetic superconductor without creating Cu-O planes
is still a puzzling question. There are still many unanswered questions
regarding the origin of magnetism and superconductivity in the Cu-substituted
Sr$_{2}$YRuO$_{6}$ compounds. At the same time, there are no detailed
magnetization studies available for the parent compound itself, except
for one report on Sr$_{2}$YRuO$_{6}$ single crystals \cite{G. Cao  (2001)}
which confirms the magnetic ordering and weak ferromagnetism. In addition,
the resistivity of Sr$_{2}$YRuO$_{6}$ single crystals \cite{G. Cao  (2001)}
shows anomalous behaviour below $T_{N}$ followed by a Mott-like transition
at 17\,K whereas the magnetoresistance becomes negative below 30\,K.
The band structure calculations \cite{E. V. Kuzmin  (2003)} have
indicated the competition between the antiferromagnetic and ferromagnetic
fluctuations among the Ru moments. We present here some additional
evidence for the competition between antiferromagnetic and ferromagnetic
coupling in this compound. Detailed measurements of magnetization
and heat capacity show some anomalous properties exhibited by Sr$_{2}$YRuO$_{6}$.
Both the measurements unfold clear evidence for two magnetic orderings
($T_{M1}\sim32$~K and $T_{M2}\sim27$~K), even though the magnetic
ordering in this compound can occur only by Ru moments. The magnetization
measurements corroborate that both the magnetic ordering occurs with
ferromagnetic components and these two components align opposite to
each other with respect to the magnetic field direction, resulting
in a magnetization reversal. The results presented here have relevance
to the magnetic properties exhibited by the Cu-substituted superconducting
Sr$_{2}$YRu$_{1-x}$Cu$_{x}$O$_{6}$ compounds \cite{Wu (2000),Marco (2000),Blackstead (2000),D. R. Harshman (2000),H. A. Blackstea (2001),Harshman  (2003)}.%
\begin{figure}
\begin{centering}
\includegraphics[scale=0.42]{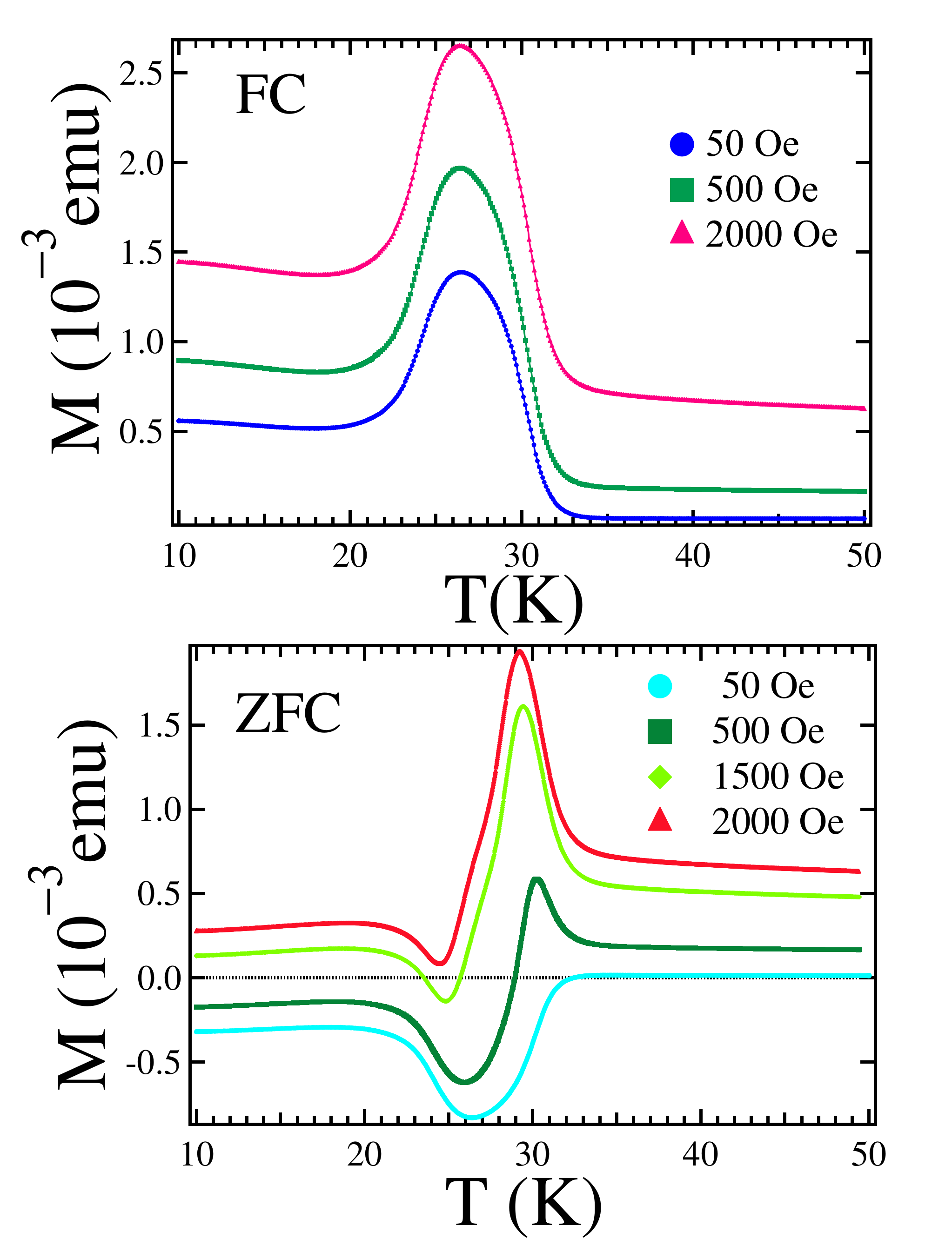}
\par\end{centering}

\caption{(Color online) Magnetization vs temperature for Sr$_{2}$YRuO$_{6}$
in zero field-cooled (ZFC) and field-cooled (FC) modes under various
applied fields.}

\end{figure}

\section{EXPERIMENTAL DETAILS}

Polycrystalline samples of Sr$_{2}$YRuO$_{6}$ were prepared by the
standard solid state reaction method by mixing stoichiometric amounts
of SrCO$_{3}$, Y$_{2}$O$_{3}$ and Ru metal powder and heating at
960$^{\circ}$C for 12 hours. The final sintering of the pelletized
powder was carried out at 1360$^{\circ}$C for 24 hours after several
intermediate heat treatments followed by grindings. X-ray diffraction
pattern of the samples was recorded on an X'pert PRO diffractometer
(PANalytical, Holland). The magnetization as a function of temperature
and magnetic field was measured using a vibrating sample magnetometer
(Quantum design, USA). The heat capacity measurements using the relaxation
method were performed using a physical property measurement system
(Quantum design, USA) in the temperature range 1.8-300~K.

\section{RESULTS AND DISCUSSION}

The Rietveld analyses of the x-ray diffraction patterns using Fullprof
software showed that the compound forms in single phase with a monoclinic
structure (space group $P2_{1}/n$). The lattice parameters obtained
from the analyses are, $a=5.769$~\AA, $b=5.772$~\AA~ and $c=8.159$~\AA~
along with $\beta=90.18^{\circ}$ which are in good agreement with
those reported earlier \cite{P. D. Battle and W. J. Macklin (1984)}.
Figure 1 illustrates the magnetization of Sr$_{2}$YRuO$_{6}$ as
a function of temperature in zero field-cooled (ZFC) and field-cooled
(FC) modes. In the ZFC measurements the sample was cooled in zero
applied field to 2~K, the required magnetic field was applied and
then the data were taken while increasing the temperature. For the
FC measurements, the sample was cooled from the paramagnetic state
to 2~K in an applied field and the data were recorded while heating
the sample. In order to minimize the remnant field in the superconducting
magnet before the ZFC measurements, the magnetic field was reduced
to zero from a large field value in the oscillating mode. This made
sure that the remnant field was within $\pm2$~Oe. The lower panel
shows the ZFC measurements for various applied field values. For low
field values, the magnetization is negative at lower temperatures.
As the temperature is increased, the magnetization remains independent
of temperature till $\sim20$~K and then surprisingly decreases to
go through a minimum. As the temperature is further increased, the
magnetization increases, goes through a positive maximum and then
shows the normal paramagnetic behaviour. For $H=1.5$~kOe, the ZFC
magnetization starts with a positive value at low temperatures, but
goes through negative value at the minimum. For higher fields, the
magnetization is always positive, even though it goes through a minimum.
The width of the peaks at the maximum and minimum as well as the temperature
at which they occur depends slightly on the applied fields; both decrease
with increasing field. The upper panel of Fig.~1 shows the FC measurements
at various applied fields. The FC magnetizations show a broad peak
and the temperature at which the peak occurs shows a weak temperature
dependence on the applied fields.%
\begin{figure}[b]
\includegraphics[scale=0.32]{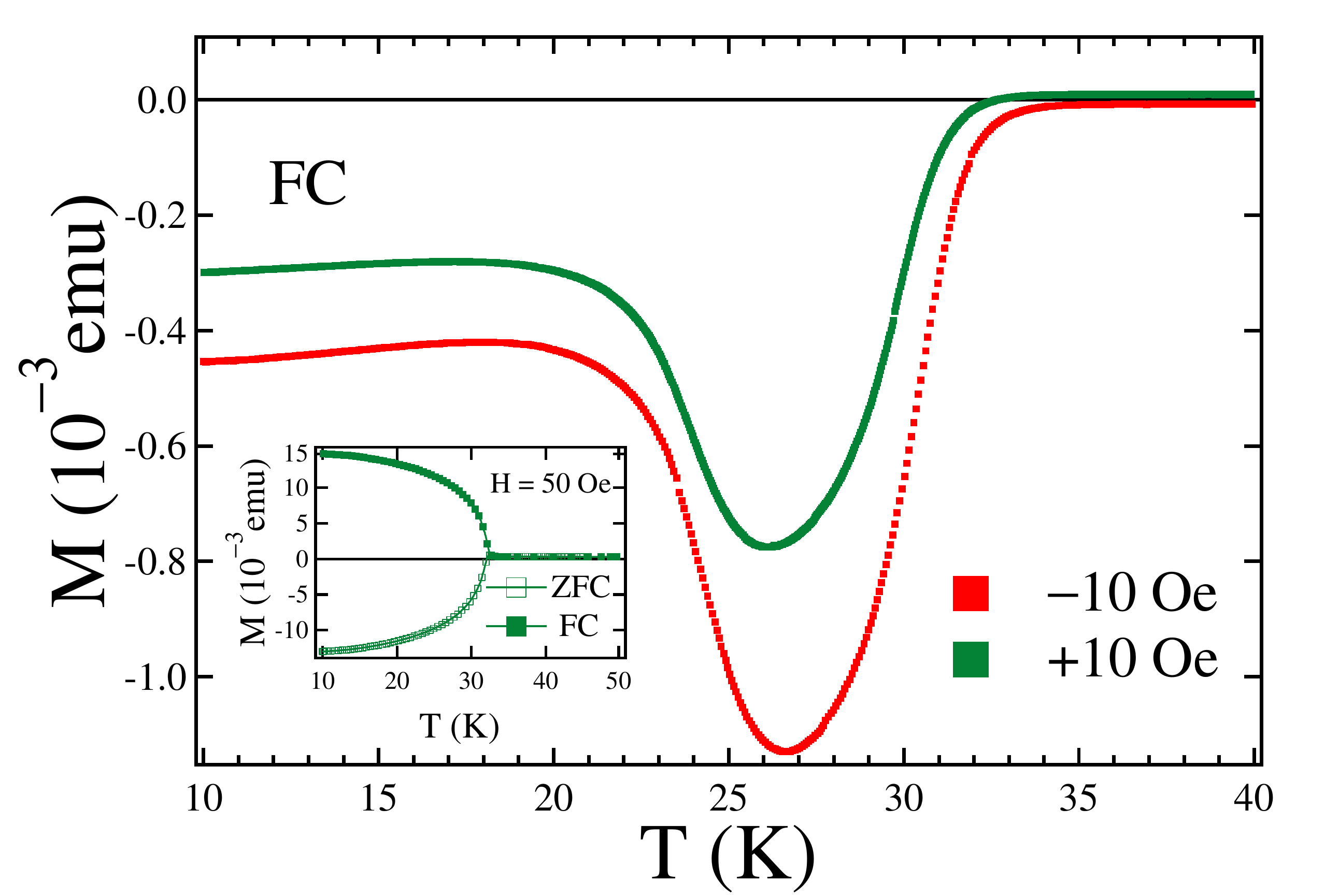}

\caption{(Color online) Magnetization for Sr$_{2}$YRuO$_{6}$ in FC mode at
$\pm10$~Oe. Inset shows the FC and ZFC curves for MnCO$_{3}$.}

\end{figure}

In order to ascertain that the anomalies observed in the ZFC magnetization
is not entirely due to effects of negative remnant magnetic field
in the superconducting coils, we have carried out FC measurements
in smaller field values, both positive and negative. Figure 2 shows
the FC measurements for an applied field of $\pm10$~Oe. It is clear
that the FC magnetization remains negative whether the field is positive
or negative. Such effects are seen upto 25~Oe above which the FC
curves switch over to the positive side. Neutron diffraction studies
at 4.2~K \cite{P. D. Battle and W. J. Macklin (1984)} had indicated
only an AFM ordering of the Ru moments. It was also proposed that
the distorted monoclinic structure can give rise to a small canting
of the Ru moments and hence a small ferromagnetic component in this
compound resulting from the D-M interactions between the antiferromagnetically
ordered Ru moments. This, however, cannot explain the observed magnetization
behaviour in this compound. A simple ferromagnetic component due to
canting can make the magnetization negative in the ZFC mode if the
remnant field is negative. But then the magnetization will monotonically
decrease as the temperature is increased and will cross over to the
positive side before completing the magnetic order. A typical example
for such a behaviour is shown in the inset of Fig.~2 for MnCO$_{3}$,
which is a well known canted antiferromagnet having D-M interactions
\cite{MnCO3  (1956)}. We have also made sure that the compound does
not contain any SrRuO$_{3}$ impurities (not detected in x-ray diffraction
patterns) by taking the ZFC and FC data for small field values in
the temperature range 100-200~K (even a small trace of SrRuO$_{3}$
impurity will give a thermal hysteresis around its ferromagnetic ordering
temperature (150-160~K) between the ZFC and FC measurements).%
\begin{figure}
\includegraphics[scale=0.35]{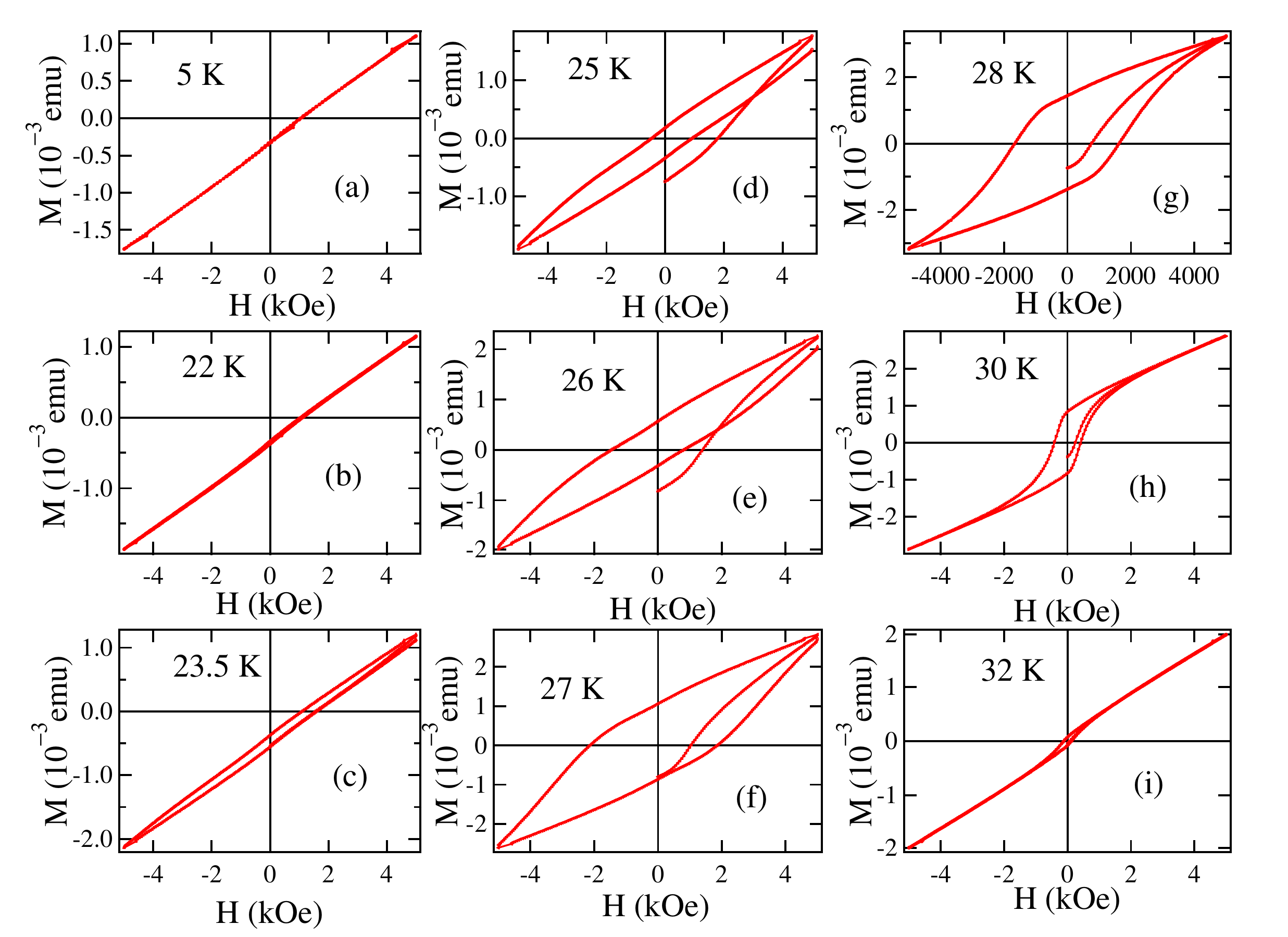}

\caption{(Color online) Isothermal magnetization curves for Sr$_{2}$YRuO$_{6}$
at different temperatures.}

\end{figure}
\begin{figure}[b]
\includegraphics[scale=0.33]{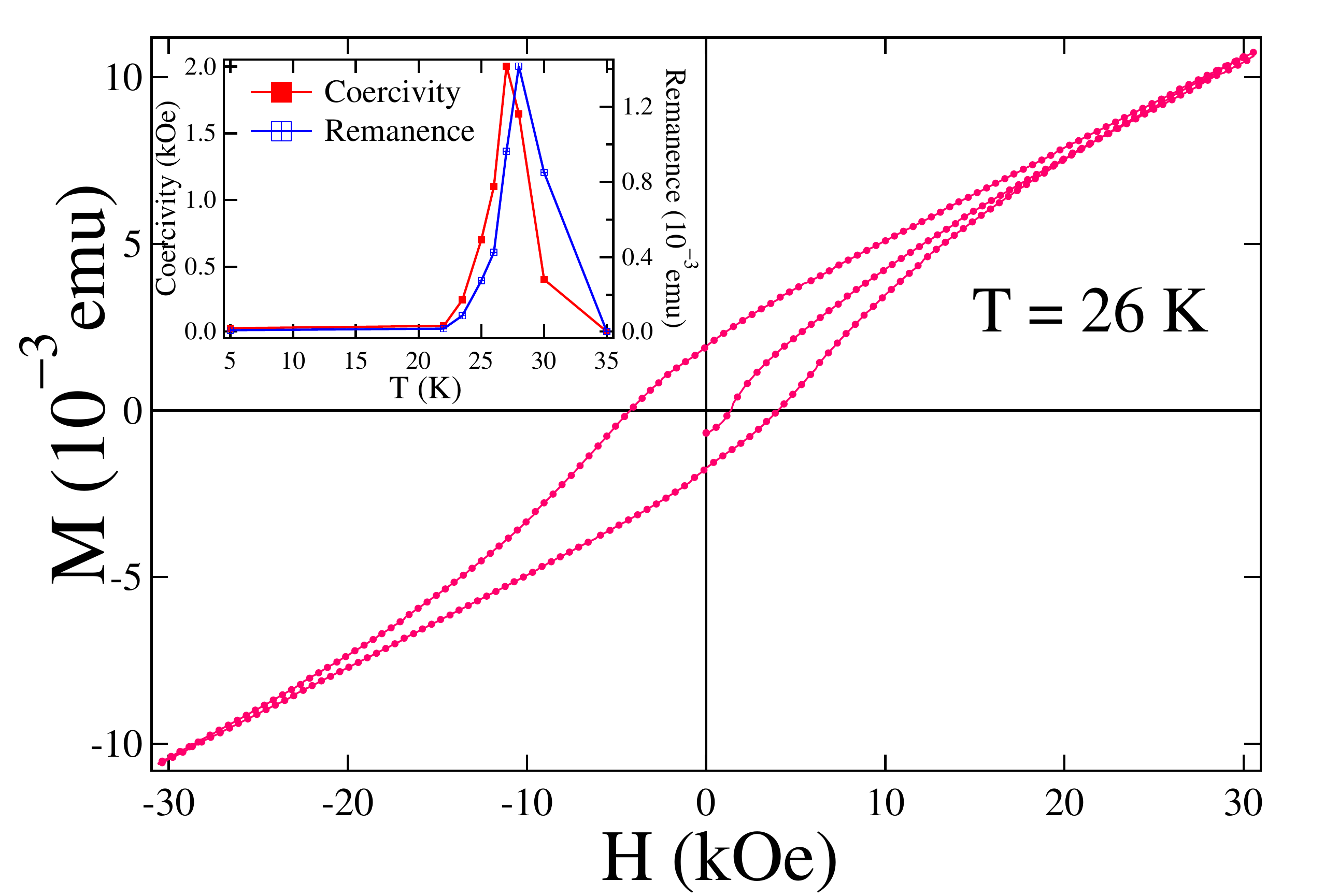}

\caption{(Color online) High field magnetization as function of field at 26~K.
Inset shows the temperature variation of coercivity ($H_{c}$) and
remnance ($M_{r}$)}

\end{figure}

In order to further corroborate that there exists more than a simple
canting in this compound, we have carried out detailed magnetization
measurements as a function of magnetic field at different temperatures.
Figure 3 depicts the low field magnetic isotherms at some selected
temperatures. At low temperatures, well below the magnetic anomalies,
the magnetization curves are almost linear with a small value for
coercivity ($H_{c}$). As the temperature crosses 22~K, the magnetization
shows significant hysteresis and the magnetization loops open up.
The opening of the loop increases until the temperature reaches $\sim27$~K
and then decreases as the temperature is further increased. Even at
32~K, the hysteresis is much more than the same at 5~K. At 35~K,
we see only a linear behaviour expected for a paramagnet. Even though
the hysteresis loops are not closed at some temperatures (Fig.~3(d),
3(e)), they show a normal behaviour when the applied fields are extended
to higher values (see main panel of Fig.~4). No other anomalies are
observed in the high field magnetization curves. The coercivity and
the remnant magnetization plotted as a function of temperature in
the inset of Fig.~4 show a maximum near 27~K and decrease on either
side of this temperature. This clearly demonstrates that some sort
of magnetization reversal happens at 27~K.

There are no reports about the heat capacity of this compound in the
literature. The result of our heat capacity measurements for Sr$_{2}$YRuO$_{6}$
is presented in Fig.~5(a). Two peaks are obvious, one at $T=\sim30$~K
and the other at $T=\sim26$~K, which correspond well to the anomalies
observed in the magnetization. There is only a minor effect by the
magnetic field on the heat capacity of the sample even at 50\,kOe
(Fig.~5(a)), even though a small decrease in the temperature dependence
of the peak positions were observed in the magnetization measurements.
In order to have an estimate of the approximate magnetic heat capacity,
the phonon contribution needs to be subtracted from the total measured
heat capacity. Since there are no nonmagnetic analogues available
for this compound, the phonon contribution was calculated from the
combined Debye and Einstein equations \cite{C. A. Martin  (1991)},

\begin{equation}
C_{{\rm ph}}=R\left(\frac{1}{1-\alpha_{D}}\left(\frac{\theta_{D}}{T}\right)^{3}\int_{0}^{x}\frac{x^{4}e^{x}}{\left(e^{x}-1\right)^{2}}dx+\sum_{i=1}^{3n-n}\frac{1}{1-\alpha_{E}}\,\frac{y^{2}e^{y}}{(e^{y}-1)^{2}}\right)\end{equation}
where $\alpha_{E}$ and $\alpha_{D}$ are the anharmonicity coefficients,
$\theta_{D}$ is the Debye temperature, $\theta_{E}$ is the Einstein
temperature, $x=\theta_{D}/T$ and $y={\theta_{E}}_{i}/T$. The best
possible fit was obtained when the calculations were performed by
using one Debye and three Einstein frequencies along with a single
$\alpha_{E}$. The solid line in Fig.~5(b) represents the fit to
the phonon contribution, which is in good agreement with the experimental
data at high temperatures (above the magnetic ordering). The parameters
obtained from the best fit are: $\theta_{D}=200$~K, $\theta_{E1}=300$~K,
$\theta_{E2}=529$~K, $\theta_{E3}=615$~K, $\alpha_{E}=1.0\times10^{-4}$~K
and $\alpha_{D}=1.0\times10^{-4}$~K. The Debye and Einstein temperatures
obtained for Sr$_{2}$YRuO$_{6}$ are comparable with those obtained
for YVO$_{3}$ where the phonon contribution was obtained in a similar
fashion, but with only two Einstein frequencies along with one Debye
frequency \cite{G. R. Blake  (2002)}. The magnetic heat capacity,
$C_{mag}$, obtained by subtracting the calculated phonon heat capacity
from the total heat capacity, is shown in the Fig. 5(c) along with
the magnetic entropy ($S_{mag}=\int_{T_{1}}^{T_{2}}\frac{C_{mag}}{T}dT$).
The two peaks become more obvious in the magnetic heat capacity. In
Sr$_{2}$YRuO$_{6}$, the magnetic transition can occur only due to
the ordering of Ru$^{5+}$ moments. In that case, the exact reason
for the observed double peak behaviour is not very clear at present.
Magnetic entropy ($S_{mag}$) increases with temperature and saturates
to a value of $\sim$~2.6~J~mole$^{-1}$~K$^{-1}$ above 30~K.
If we consider the ground state of Ru$^{5+}$ ions as $J=3/2$, then
the expected magnetic entropy is 11.52~J~mole$^{-1}$~K$^{-1}$
($S=R\ln[2J+1]$), corresponding to the four-fold degenerate ground
sate. However, the crystalline electric fields, if present, can split
this ground state into two doubly degenerate states, giving rise to
a ground state multiplicity of only two \cite{Y. Doi  (2003)}. This
will reduce the magnetic entropy of the compound to 5.76~J~mole$^{-1}$~K$^{-1}$
($S=R\ln2$). The observed entropy, however, is even less than half
of this value. In fact, neutron diffraction measurements had estimated
a value of 1.8~$\mu_{B}$/Ru$^{5+}$ at 4~K (instead of the expected
value of 3~$\mu_{B}$/Ru$^{5+}$) in the magnetically ordered state
\cite{P. D. Battle and W. J. Macklin (1984)}. This moment value corresponds
well with the doubly degenerate ground state. If we compare the reduction
in entropy of Sr$_{2}$YRuO$_{6}$ to that observed for YVO$_{3}$
\cite{G. R. Blake  (2002)}, frustrations of Ru spins at high temperatures
(above the magnetic ordering) can be attributed as the reason for
the reduction in entropy. The correlation between the frustrated moments
at high temperatures reduces the contribution of the entropy to the
magnetic ordering. Such a frustration among the Ru moments is inferred
in Sr$_{2}$YRuO$_{6}$ as the possible reason for the reduction in
$T_{N}$ even though the compound possesses a large exchange integral
value \cite{E. V. Kuzmin  (2003)}.%
\begin{figure}
\includegraphics[scale=0.5]{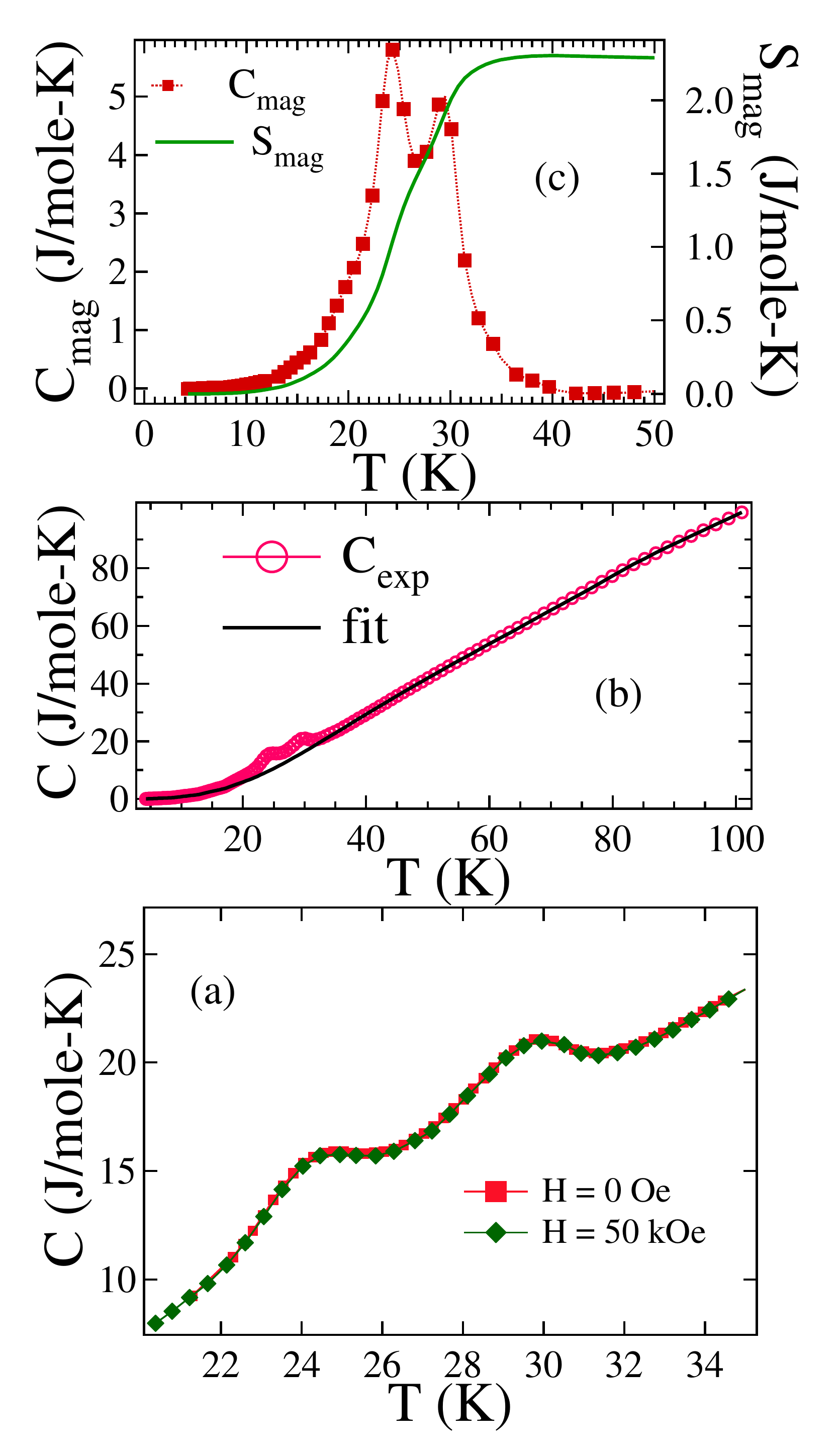}

\caption{(Color online) (a) Measured total heat capacity ($C_{v}$) of Sr$_{2}$YRuO$_{6}$
in applied fields $H=0$~Oe and 5~kOe. (b) Heat capacity with the
fit (solid line) for phonon contribution. (c) Magnetic contribution
to heat capacity ($C_{mag}$) (left scale) and calculated magnetic
entropy ($S_{mag}$) (right scale) as a function of temperature.}

\end{figure}

It is clear that Sr$_{2}$YRuO$_{6}$ exhibits two anomalies, the
first at $\sim32$~K ($T_{M1}$) and the second at $\sim27$~K ($T_{M2}$)
even though the magnetic ordering in this compound can come only from
the Ru$^{5+}$ moments. If we assume that the two anomalies are associated
with the magnetic ordering of the Ru moments, then the observed behaviour
is very interesting. The isothermal magnetization curves at different
temperatures (Fig.~3) clearly demonstrate that the first magnetic
order starts at $T_{M1}\sim32$~K with a ferromagnetic component
resulting in the increase of hysteresis and $H_{c}$ as the temperature
is lowered. This ferromagnetic component is expected from the canting
of the antiferromagnetically ordered Ru moments because of the D-M
interactions. However, the decrease of hysteresis and $H_{c}$ below
27~K indicates that a second component of the magnetic order also
develops with a ferromagnetic component ($T_{M2}$), but aligns itself
opposite to the first component and hence opposite to the applied
field. This component almost cancels the first component and hence
the hysteresis is negligible at low temperatures ($<20$~K). These
anomalies are further confirmed in the zero field remnant magnetization
measurements as shown in Fig.~6(a). Here the sample was cooled (FC)
in a field of 5~kOe down to 10~K. The field was then removed and
the remnant magnetization was measured in zero field while warming
the sample. The remnant magnetization shows a normal decrease upto
$\sim20$~K, but then increases, goes through a maximum at $\sim27$~K
and then decreases to zero above 32~K. This clearly demonstrates
that the magnetic ordering consists of two components and they are
aligned opposite to each other with respect to the magnetic field
direction. While cooling the sample in magnetic field, the first component
orders and aligns parallel to the field at $\sim32$~K, but the second
component aligns antiparallel to the field at $\sim27$~K, decreasing
the net magnetization. However, this antiparallel component is not
strong enough to make the magnetization negative as in the case of
some $Ln$VO$_{3}$ compounds. As the sample is warmed up in zero
field, the remnant magnetization increases when the antiparallel component
relaxes and completes its disordering. %
\begin{figure}
\includegraphics[scale=0.5]{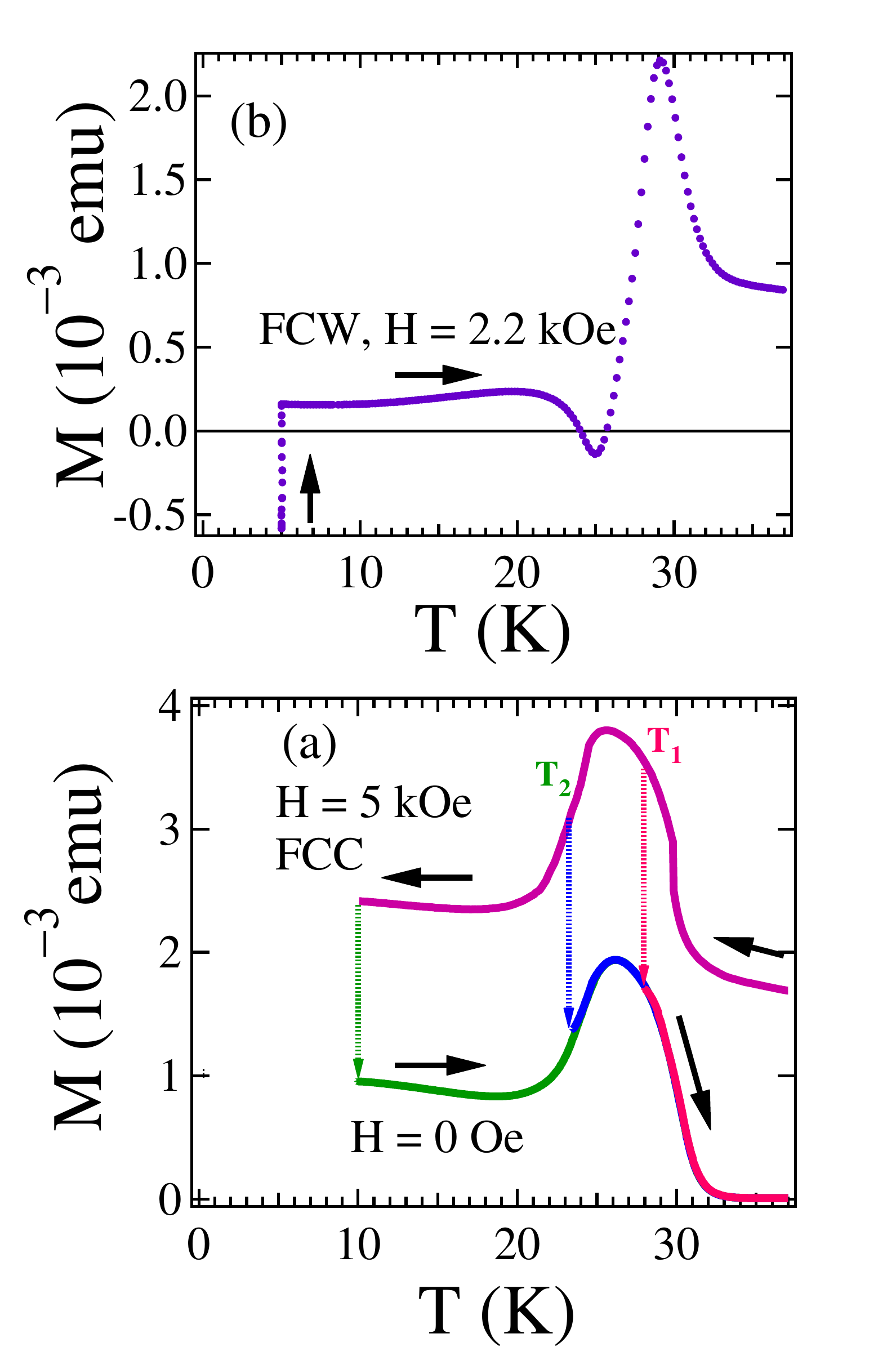}

\caption{(Color online) (a) FC data while cooling the sample in an applied
field of 5\,kOe (top curve) and the zero field warm up data (bottom
curves) after switching off the applied magnetic field at $T$ = 10\,K,
23.5\,K and 28\,K. (b) FC warm up data in an applied field of 2.2\,kOe
after cooling the sample in a negative applied field of $-$50\,Oe.}

\end{figure}

In order to verify the thermodynamic reversibility of the FC state,
we have measured the remnant magnetization by field-cooling the sample
down to two pre-selected temperatures ($T_{1}$ and $T_{2}$) on either
side of the maximum in the FC magnetization curve as shown in Fig.~6(a).
In the first case, the sample was field cooled to 28~K ($T_{M2}<T_{1}<T_{M1}$),
where only the first magnetic component would have ordered, and the
field was removed. In the second case, the sample was field cooled
to 23.5~K ($T_{2}<T_{M1},T_{M2}$), where both components would have
undergone magnetic ordering, before removing the field. In both cases,
the remnant magnetization follows exactly the original remnant curve
which was obtained by switching the field off at 10~K, proving the
thermodynamic reversibility of the magnetically ordered states. Further
evidence for the opposite alignments of the two components of magnetic
ordering is evident from the data in Fig.~6(b). Here the sample was
cooled in a negative field ($-50$~Oe) so that the magnetization
at 5~K is negative. At 5~K, the field was increased in the positive
direction ($\sim2.2$~kOe) until the magnetization became positive.
Magnetization shows normal behaviour upto 20~K, but shows a sudden
dip to go through a negative minimum at $\sim$27~K. Thus it is clear
that whether the sample is cooled in positive or negative field, the
two components of magnetic order align always opposite to each other.

Neutron diffraction measurements in this compound at 4.2~K\textbf{
}\cite{P. D. Battle and W. J. Macklin (1984)} had indicated that
the nuclear structure of the compound remains unchanged at 4.2~K,
ruling out the possibility of any structural changes. On the other
hand if we assume that one of the anomalies is due to a possible structural
change, then the changed structure should again revert back to the
original structure at low temperatures. From the position of the magnetic
diffraction peaks, magnetic ordering of the Ru moments was deduced
to be antiferromagnetic with a type~I structure \cite{P. D. Battle and W. J. Macklin (1984)}.
However, the temperature variation of the intensity of magnetic peak(s)
was not reported and hence the exact temperature at which the Ru moments
order is not available. The $T_{N}$ of $\sim26$~K was assigned
to this compound simply from the position of the peak in the high
field magnetization measurements \cite{P. D. Battle and W. J. Macklin (1984)}.
Even though there are no detailed experimental observations in Sr$_{2}$YRuO$_{6}$,
many experimental results including neutron diffraction exist for
Cu-substituted Sr$_{2}$YRu$_{1-x}$Cu$_{x}$O$_{6}$ compounds\textbf{
}\cite{Blackstead (2000),H. A. Blackstea (2001),Harshman  (2003),Marco (2000)}.
Assuming that the magnetic properties associated with the ordering
of the Ru moments are not drastically altered, we can analyze the
results of the $\mu$SR measurements in the Cu-substituted compounds
\cite{Harshman  (2003),Blackstead (2000)}. Both the precession frequency
and relaxation rate show anomalies at $\sim30$~K for muons trapped
in the two possible sites; oxygen in the YRuO$_{4}$ layers ($\mu_{O(1,2)}$)
and oxygen in the SrO layers ($\mu_{O(3)}$). The authors assigned
this anomaly to the fluctuation of the Ru moments which order in a
spin-glass state. This is exactly the same temperature range at which
the first component of magnetic ordering starts in our studies for
the parent compound ($T_{M1}$). The variation of the power exponent
at the $\mu_{O(1,2)}$ site as a function of temperature \cite{Harshman  (2003)}
almost resembles the mirror image of our temperature variation of
the coercivity and remnant magnetization (inset of Fig.~4). It was
also proposed that the fluctuations of the Ru moments stop at $T=23$~K
and order antiferromagnetically. We have identified this temperature
in our studies as the temperature at which the second component of
magnetic moments completes its ordering. $^{\text{99}}$Ru M\"ossbauer
measurements in Sr$_{2}$YRu$_{1-x}$Cu$_{x}$O$_{6}$ ($x=0.05$)
\cite{Blackstead (2000),Marco (2000)} have confirmed the magnetic
ordering of Ru moments. Both the isomer shift and the hyperfine field
values are consistent with the pentavalent nature of the Ru moments.
From the temperature variation of the M\"ossbauer spectra \cite{Marco (2000)},
it was concluded that that the magnetic ordering of the Ru moments
persists upto 30~K, which is consistent with the magnetic ordering
temperature $T_{M1}$ in our measurements. Additional support for
the Ru ordering as high as 30\,K comes from the neutron diffraction
measurements reported for Sr$_{2}$YRu$_{1-x}$Cu$_{x}$O$_{6}$ ($x=0.15$)
where the intensity of the magnetic peak due to the ordering is Ru
moments is evident upto $\sim30$~K \cite{H. A. Blackstea (2001)}.

In ZFC measurements, the sample is always cooled in a nominal remnant
field ($\sim\pm2$~Oe in our case). This may result in the negative
magnetization at low temperatures. When the magnetic field is applied,
the magnetization remains negative for low fields (Fig.~1) but changes
over to positive values for sufficiently large fields. Even then a
magnetization reversal happens and the magnetization goes though a
negative minimum in between the two transitions. This behaviour may
be caused by the fact that the large fields flip (to orient parallel
to the field) some, but not all, of the spins oriented against the
field. As the field value increases, majority of the spins, oriented
antiparallel to the field, align parallel to the field, resulting
only in a small minimum at $T_{M2}$. In FC magnetization ($H\geq50$~Oe),
the first component of the magnetic transition aligns along the field
at $T_{M1}$, resulting in a positive magnetization. However, at $T_{M2}$,
the second component aligns antiparallel to the first component, resulting
in the reduction in magnetization. When this alignment is over, the
magnetization remains constant, as in the case of FC measurements
in smaller fields.

The reason for the two components of the magnetic ordering reminiscent
of ferrimagnetic ordering in this compound is not clear. In the ordered
double perovskite Sr$_{2}Ln$RuO$_{6}$ compounds, the $B$-site of
the perovskite structure ABO$_{3}$ is uniquely occupied either by
Ru or the rare earth metal ions due to the lower coordination number
compared to Sr. Since Ruthenium is considered to be in the oxidation
state of 5+ in these compounds, the chances of ferrimagnetic ordering
of Ru$^{5+}$/Ru$^{4+}$ moments are very rare. We have repeated the
measurements with samples annealed in air, oxygen or argon and all
of them showed the same behaviour. In fact, the isomer shift values
from the $^{99}$Ru M\"ossbauer measurements in Sr$_{2}$YRuO$_{6}$
had confirmed the pentavalent oxidation state of Ru moments \cite{Blackstead (2000),Greatrex  (1979)}.
Even if the spin-glass state is assumed as suggested by Harshman et
al \cite{Harshman  (2003)}, the observed properties - the magnetization
reversal, thermodynamic reversibility of magnetization and two peaks
in heat capacity - cannot be explained. The magnetic interactions
among the ordered Ru moments can take place in two ways; (i) the $\sigma$-super
exchange interaction between nearest neighbour (nn) Ru$^{5+}$ ions
via Ru-O-O-Ru pathway and (ii) the $\pi$-super exchange between the
next nearest neighbour (nnn) Ru$^{5+}$ ions via Ru-O-Y-O-Ru pathway.
The relative strengths of these two interactions will determine the
type of magnetic ordering at low temperatures. Since Y is a nonmagnetic
ion, it is not expected to take part in the exchange interaction and
hence the second interaction between the nnn is assumed to be negligible.
This assumption can further be supported by the fact that the relaxation
rate at the $\mu_{O(3)}$ site for the muons is very much smaller
than the relaxation rate at the $\mu_{O(1,2)}$ site \cite{Harshman  (2003),Blackstead (2000)}
since the Ru-O-Y-O-Ru pathway includes the oxygen at the O(3) sites
only. Whether the competition between these two interactions, however,
can give rise to the observed double peak behaviour needs further
investigations. It is possible that the stronger interaction orders
the Ru moments at $T_{M1}$ and the weaker interaction realigns some
of the Ru moments in the opposite direction at $T_{M2}$. It is possible
that the second component of the magnetic order at $T_{M2}$ is not
strong enough to bring the magnetization down to negative values even
though it gets reduced. However, why the second component aligns the
moments against the first component and hence the magnetic field is
not clear now. Detailed neutron diffraction measurements are needed
as a function of temperature to explain the reason for the observed
anomalies.

In conclusion, we have reported some of the anomalous properties exhibited
by SrYRuO$_{6}$ deduced from the detailed magnetic and heat capacity
measurements. Two distinct magnetic orderings are evident in the compound
even though only Ru moments can order in this compound. The two components
of the magnetic order align always opposite to each other and to the
magnetic field. Observation of hysteresis and coercivity in magnetic
isotherms below the magnetic order indicates the presence of ferromagnetic
component associated with the magnetic ordering. The presence of two
well defined peaks in the heat capacity indicates that the two magnetic
components have large entropy change associated with the ordering.

\end{document}